\documentclass[runningheads]{llncs}

\usepackage[T1]{fontenc}
\usepackage{graphicx}
\usepackage{booktabs}
\usepackage{multirow}
\usepackage{colortbl}
\usepackage{xcolor}
\usepackage{amsmath}
\usepackage{amssymb}
\usepackage{float}
\usepackage[hidelinks]{hyperref}

\begin{document}

\title{A Two-Stage Multi-Modal MRI Framework for Lifespan Brain Age Prediction}

\titlerunning{Multi-Modal Lifespan Brain Age Prediction}

\author{Dingyi Zhang\inst{1}\thanks{Equal contribution.} \and
Ruiying Liu\inst{2}$^\star$ \and
Yun Wang\inst{1,2}}

\authorrunning{D. Zhang and R. Liu et al.}

\institute{Department of Computer Science, Emory University, Atlanta, GA, USA \and
Department of Biomedical Informatics, Emory University, Atlanta, GA, USA \\
\email{dingyi.zhang@emory.edu, rliu60@emory.edu, yun.wang2@emory.edu}}

\maketitle

\begin{abstract}
The accurate quantification of brain age from MRI has emerged as an important biomarker of brain health. However, existing approaches are often restricted to narrow age ranges and single-modality MRI data, limiting their capacity to capture the coordinated macro- and microstructural changes that unfold across the human lifespan. To address these limitations, we develop a multi-modal brain age framework to characterize the integrated evolution of brain morphology and white matter organization. Our model adopts a two-stage architecture, where modalities are processed independently and integrated via late fusion in both stages: first to estimate a probability distribution over six developmental stages, and then to predict age via probability-weighted stage-specialized experts. Experiments on nine datasets spanning fetal to elderly stages demonstrate competitive in-domain performance and out-of-domain generalization, with our method reducing MAE by 13\% and 78\% over existing baselines and multi-modal integration yielding 12--13\% gains. Analysis of ADNI clinical groups further suggests the potential of the predicted brain age gap to characterize Alzheimer's-related brain aging.
\end{abstract}


\section{Introduction}

Brain age prediction---the task of estimating the biological age of the brain from MRI---has emerged as a principled way to quantify an individual's neuroanatomical development and aging~\cite{kumari2024review}. The gap between predicted brain age and chronological age, known as the brain age gap (BAG), is a sensitive biomarker: accelerated brain aging is associated with Alzheimer's disease, schizophrenia, and preterm birth complications, while decelerated aging may reflect protective factors~\cite{baecker2021machine}.

With the growth of large-scale neuroimaging datasets and advances in deep learning, data-driven methods~\cite{franke2010estimating,cole2017predicting} have achieved impressive accuracy on specific cohorts. However, four fundamental limitations restrict the applicability of existing methods.

\textbf{Limited age coverage.} Most brain age models are trained on a single developmental window and fail to generalize outside it. The brain's morphological appearance changes so dramatically across the fetal, infant, child, adolescent, adult, and elderly periods~\cite{bethlehem2022brain} that a model trained on one stage produces substantially degraded predictions on another.

\textbf{Single-modality design.} Most methods rely on a single MRI modality, capturing only one aspect of brain structure. T1-weighted (T1w) MRI reflects gross morphology such as cortical thickness and tissue volumes, T2-weighted (T2w) MRI provides complementary tissue contrast particularly important during early neurodevelopment, and diffusion-derived fractional anisotropy (FA) characterizes white matter microstructure and maturation~\cite{kumari2024review}.

\textbf{Missing modality fragility.} Multi-site datasets routinely lack one or more imaging sequences due to scanner or protocol differences. Multi-modal approaches that require all modalities at inference cannot be directly applied under such conditions, limiting real-world deployability.

\textbf{Limited evaluation across heterogeneous and clinical populations.} Most existing brain age prediction methods are evaluated on relatively restricted cohorts, often using training and test data drawn from the same dataset or from populations with similar acquisition protocols. Such evaluation settings may not fully reflect the heterogeneity of real-world neuroimaging data, where scans are acquired across different sites, scanners, imaging protocols, age ranges, and clinical conditions. In particular, external validation on out-of-domain datasets and disease cohorts remains limited, making it difficult to assess whether learned age-related representations generalize beyond the training distribution. This limitation is especially important for lifespan brain age prediction, where robust models must account for both acquisition-related domain shifts and biological variability across healthy and pathological development~\cite{fortin2018harmonization,wachinger2021detect,dinsdale2021learning}.

To address these limitations, we propose a two-stage multi-modal framework for lifespan brain age prediction. The lifespan is highly heterogeneous for direct regression, as cortical folding, myelination, synaptic pruning, and atrophy unfold across vastly different timescales. We therefore adopt a coarse-to-fine hierarchy: Stage~I estimates a probability distribution over developmental stages and Stage~II performs fine-grained regression via probability-weighted stage experts. Both stages share a 3D CNN backbone pretrained with a masked autoencoding objective~\cite{he2022masked}, equipped with a Mixture of Experts (MoE)~\cite{shazeer2017outrageously} that routes each modality to its designated expert. Late fusion ensures robustness to any subset of available modalities without retraining. Our main contributions are as follows:
\begin{enumerate}
    \item We present a unified brain age prediction framework spanning the complete human lifespan, from fetal (gestational week 20) to elderly stages (age 100+), within a single model.
    \item We design a two-stage coarse-to-fine pipeline where Stage~I identifies the developmental period and Stage~II estimates continuous age within it; the shared encoder is equipped with a Modality MoE for modality-specific feature extraction, and Stage~II further employs an Age Stage MoE trained with teacher forcing and soft-routed at inference via Stage~I probability weighting.
    \item We adopt a late fusion strategy integrating T1w, T2w, and FA modalities that requires no multi-modal training pairs and accommodates missing modalities at inference.
    \item We conduct extensive evaluation on four out-of-domain datasets demonstrating improved cross-site and cross-scanner generalization, and further validate our model on ADNI clinical populations, suggesting its potential to characterize Alzheimer's-related brain aging.
\end{enumerate}

\section{Related Work}

\subsection{Brain Age Prediction}

Brain age prediction has progressed from classical morphometric regression to deep learning.
Franke et al.~\cite{franke2010estimating} established an early benchmark through support vector regression; Cole et al.~\cite{cole2017predicting} demonstrated that 3D CNNs outperform classical pipelines and formalized BAG as a heritable biomarker associated with Alzheimer's disease and psychiatric conditions.
Subsequent methods improved accuracy through architectural and loss design innovations: SFCN~\cite{peng2021accurate} treats age as soft classification over discretized bins; TSAN~\cite{cheng2021brain} employs a coarse-to-fine cascade with ranking loss; and ORDER~\cite{shah2024ordinal} mitigates regression-to-the-mean bias via ordinal classification with distance regularization.
Beyond single-modal architectures, Dinsdale et al.~\cite{dinsdale2021learning} showed that combining T1w and diffusion MRI reveals complementary aging signatures; subsequent multimodal frameworks confirmed consistent gains across diverse cohorts and modality combinations~\cite{liem2017predicting,de2020multimodal,guan2024brain}.

\subsection{Neuroimaging Modalities Across Development}

Brain age prediction across the full lifespan remains challenging because brain anatomy, tissue composition, and MRI contrast vary substantially with development and aging. However, most existing brain age prediction models are designed for a single imaging modality, most commonly T1-weighted MRI, and are often restricted to a specific developmental or aging window~\cite{peng2021accurate,cheng2021brain,shah2024ordinal}. This limits their ability to capture age-related changes that are expressed differently across imaging contrasts and life stages. During fetal and neonatal development, rapid tissue growth, immature myelination, and high water content produce image contrasts that differ markedly from those observed in the adult brain. In this period, T2-weighted MRI is commonly preferred because it provides more reliable anatomical contrast for the developing brain, whereas T1-weighted contrast is less informative before substantial myelination has occurred~\cite{payette2021automatic}. As myelination progresses through infancy and childhood, T1-weighted MRI becomes increasingly useful for quantifying macroscopic morphological features, including cortical thickness, tissue volumes, and subcortical structures. Diffusion MRI provides further complementary information by characterizing white matter microstructure. Measures such as fractional anisotropy and mean diffusivity capture age-related changes in tissue organization, including white matter maturation during development and microstructural degeneration during aging~\cite{ball2014rich,taoudi2022predicting,bethlehem2022brain}. Therefore, T1-weighted, T2-weighted, and diffusion MRI provide distinct but complementary information for modeling brain age across the lifespan. Nevertheless, the age-related information contained in T2-weighted and diffusion MRI remains underexplored in existing brain age prediction frameworks, particularly for fetal, neonatal, and early postnatal development.

\subsection{Self-Supervised Representation Learning for Lifespan MRI}

Although chronological age is typically available for brain age prediction, constructing robust models across the full lifespan remains challenging because data are highly imbalanced across developmental stages and imaging modalities. Adult cohorts are substantially more abundant than fetal, neonatal, and early postnatal cohorts, and age definitions in early development may vary across studies, including gestational age, postmenstrual age, corrected age, or scan age. In addition, MRI contrast and modality availability differ across the lifespan, making supervised training on limited early-development data particularly unstable.

Self-supervised learning provides a way to learn generalizable image representations from large-scale unlabeled or weakly labeled neuroimaging data before task-specific age prediction. Masked autoencoders (MAE) learn representations by reconstructing randomly masked image patches, encouraging the encoder to capture global anatomical structure rather than relying only on local intensity patterns~\cite{he2022masked}. This property is especially useful for lifespan MRI, where anatomical appearance changes substantially with age and differs across T1-weighted, T2-weighted, and diffusion-derived images. Thus, self-supervised pretraining is not used simply to compensate for missing age labels, but to improve representation learning under lifespan data imbalance, modality heterogeneity, and limited early-development samples~\cite{chen2023masked}.
\section{Method}

\subsection{Problem Formulation}

Given a set of MRI scans $\{x_m\}_{m \in \mathcal{M}}$ from available modalities $\mathcal{M} \subseteq \{\text{T1w},\allowbreak\text{T2w},\allowbreak\text{FA}\}$ preprocessed via N4 bias correction, skull stripping~\cite{hoopes2022synthstrip}, and resampled to 1~mm isotropic resolution at a spatial size of $96^3$ voxels, our goal is to predict the brain age $\hat{y} \in \mathbb{R}$ of a subject whose chronological age $y$ spans from gestational week 20 to over 100 years. The modality set $\mathcal{M}$ is subject-specific: not all sequences are available for every subject due to differing acquisition protocols, and the model must operate on any non-empty subset at both training and inference. To unify age representation across the full lifespan, fetal gestational age $w$ (in weeks) is converted to postnatal-equivalent years as $y = (w-40)/52$, placing prenatal subjects on the same continuous axis as postnatal ages.

\begin{figure}[t]
    \centering
    \includegraphics[width=\linewidth]{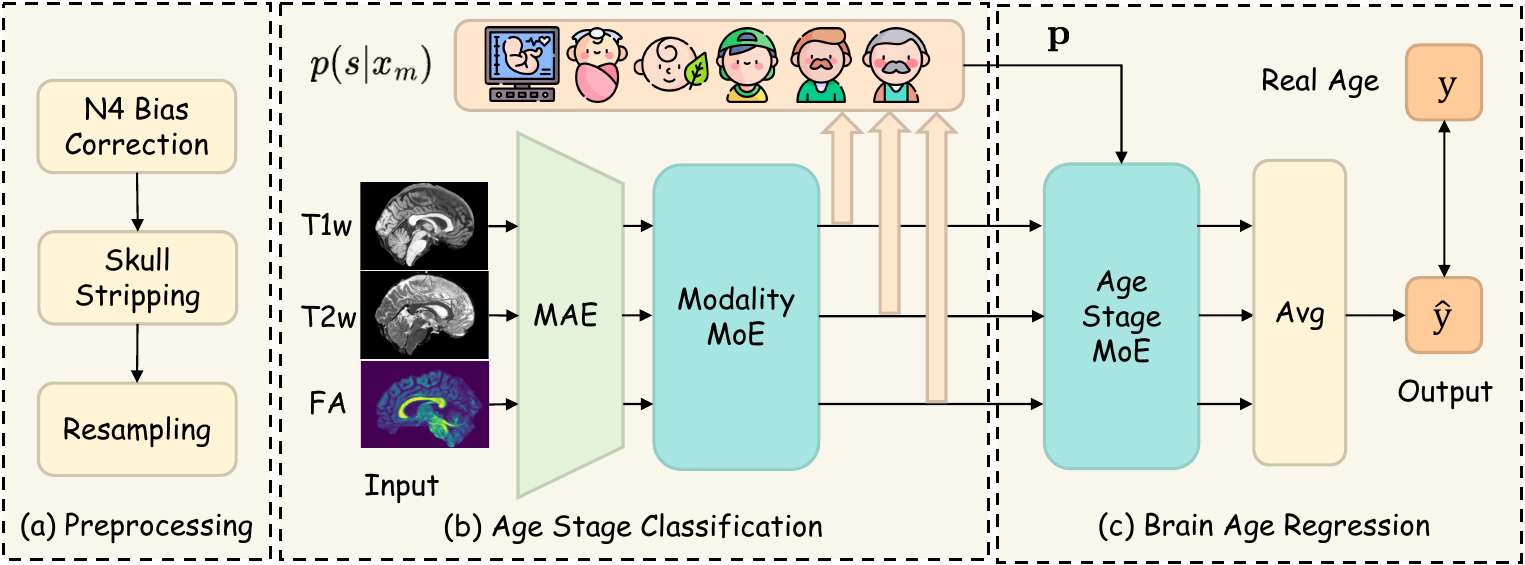}
    \caption{Two-stage multi-modal framework. Modalities are encoded independently via a shared backbone and Modality MoE. \textbf{Stage~I} maps multi-modal images to a probability distribution $\mathbf{p}$ over developmental stages; \textbf{Stage~II} uses $\mathbf{p}$ to soft-weight six stage-specialized regression experts, yielding a continuous age prediction $\hat{y}$.}
    \label{fig:framework}
\end{figure}

\subsection{Framework Overview}

Figure~\ref{fig:framework} illustrates the proposed two-stage pipeline. The full lifespan spans vastly different structural regimes, making direct regression across all ages challenging. We therefore adopt a coarse-to-fine hierarchy: Stage~I identifies the developmental period, and Stage~II performs continuous regression within it using a stage-specialized expert. Each modality is processed independently through a shared backbone, with multi-modal information combined only at the prediction level via late fusion. In Stage~I, per-modality logits are averaged to produce a probability distribution $\mathbf{p}$ over developmental stages. In Stage~II, $\mathbf{p}$ soft-weights each expert's prediction, and per-modality outputs are averaged to yield $\hat{y}$. Any non-empty modality subset can be used at inference without retraining.

\subsection{3D CNN Encoder with Modality MoE}

\paragraph{Encoder.}
Each modality volume $x_m \in \mathbb{R}^{1 \times 96^3}$ is independently encoded by a shared 3D convolutional backbone with two residual downsampling blocks. The first block expands $1 \to 256$ channels at stride~2, reducing spatial resolution from $96^3$ to $48^3$. The second block maps $256 \to 512$ channels at stride~2, yielding a dense latent feature map $f_m \in \mathbb{R}^{512 \times 24^3}$. Each block is built from ExtResNet units~\cite{milletari2016v} with GroupNorm normalization~\cite{wu2018group}, which ensures stable training under the small per-GPU batch sizes typical of 3D neuroimaging. The backbone is shared across all three modalities and both stages, providing a common low-level feature space before modality-specific MoE processing.

\paragraph{Modality-Aware MoE.}
To capture distinct structural signatures of each MRI contrast without conflating T1w morphology, T2w tissue contrast, and FA white matter microstructure, we equip the encoder with three modality-specific expert branches under a hard-routing MoE design~\cite{shazeer2017outrageously}. Each expert consists of two residual blocks followed by a channel attention layer for feature refinement. Routing is identity-based: each modality is deterministically assigned to its designated expert with no learned or data-dependent gating, ensuring each expert specialises exclusively in its assigned contrast. The output is governed by a fixed modality routing matrix $\mathbf{R} \in \mathbb{R}^{3 \times 3}$ set to the identity and kept frozen throughout training:
\begin{equation}
    f_m^{\mathrm{moe}} = \sum_{k=1}^{3} R_{mk}\, e_k(f_m),
    \label{eq:moe}
\end{equation}
where $e_k(\cdot)$ is the $k$-th expert and $R_{mk} = \mathbf{1}[m=k]$ ensures each modality is routed exclusively to its designated expert.

\paragraph{MAE Pretraining.}
Following~\cite{zhang2024mapseg}, we pretrain the encoder with a masked autoencoding objective~\cite{he2022masked} adapted for 3D convolutional representations. At each iteration, $75\%$ of spatial patches (patch size $p=4$) are randomly masked, and a convolutional decoder reconstructs the original voxel intensities at the masked locations. We pretrain on both full-resolution volumes and local high-resolution crops, providing the encoder with global anatomical context and fine-grained structural detail. The reconstruction loss is restricted to masked patches:
\begin{equation}
    \mathcal{L}_{\mathrm{MAE}} = \frac{1}{|\mathcal{P}_{\mathrm{mask}}|} \sum_{i \in \mathcal{P}_{\mathrm{mask}}} \bigl(\hat{x}_i - x_i\bigr)^2,
    \label{eq:mae}
\end{equation}
where $\mathcal{P}_{\mathrm{mask}}$ is the set of masked patch indices. This forces the encoder to learn globally consistent representations rather than relying on local texture cues, which is especially valuable for rare lifespan stages where labeled data are scarce. Full details of the pretraining datasets are provided in Supplementary Section~A.

\subsection{Stage~I: Age Stage Classification}

We partition the human lifespan into six developmental stages: \emph{fetal} (gestational weeks 20--40), \emph{neonatal} (postnatal weeks 0--12), \emph{infant} (months 3--24), \emph{child} (years 2--18), \emph{adult} (years 18--65), and \emph{elderly} (years 65--100). These boundaries align with major neurodevelopmental transitions, including rapid cortical folding, myelination onset, synaptic pruning, and late-life atrophy~\cite{bethlehem2022brain,payette2021automatic,ball2014rich}. For each modality $m$, the pretrained encoder produces $f_m^{\mathrm{moe}}$, projected to 64 channels by a connector. A Modified ResNet-18 head with four residual stages (64$\to$128$\to$256$\to$512 channels), global average pooling, and a linear classifier produces per-modality logits $\ell_m \in \mathbb{R}^6$. Late fusion averages logits across all available modalities and applies softmax to obtain a stage probability distribution:
\begin{equation}
    \mathbf{p} = \mathrm{softmax}\!\left(\frac{1}{|\mathcal{M}|}\sum_{m \in \mathcal{M}} \ell_m\right) \in \mathbb{R}^6,
    \label{eq:stage}
\end{equation}
The resulting probability distribution $\mathbf{p}$ is then used for probability-weighted expert routing in Stage II. The network is trained with cross-entropy loss:
\begin{equation}
    \mathcal{L}_{\mathrm{cls}} = \mathrm{CE}\!\left(\frac{1}{|\mathcal{M}|}\sum_{m \in \mathcal{M}} \ell_m,\; s\right),
    \label{eq:cls}
\end{equation}
where $s$ is the ground-truth. Absent modalities contribute zero logits, so the late fusion scheme handles missing modalities without any special treatment.

\subsection{Stage~II: Stage-Conditioned Age Regression}

Stage~II performs fine-grained continuous age regression within the developmental period identified by Stage~I. A separate Stage~II connector (512$\to$64 channels) and the Age Stage MoE are introduced for regression; the encoder is jointly fine-tuned at a reduced learning rate ($0.01\times$ the base rate) to preserve Stage~I lifespan representations. During training, ground-truth stage labels hard-route each sample to its designated expert (teacher forcing). Each stage expert $s$ outputs a normalized prediction $\tilde{y}_{m,s} \in [0,1]$, linearly denormalized to its stage-specific age range:
\begin{equation}
    \hat{y}_{m,s} = \tilde{y}_{m,s}\cdot\bigl(y^{(s)}_{\max} - y^{(s)}_{\min}\bigr) + y^{(s)}_{\min}.
    \label{eq:denorm}
\end{equation}
At inference, Stage~I produces a probability distribution $\mathbf{p} = \mathrm{softmax}(\hat{\mathbf{l}}) \in \mathbb{R}^6$; all six experts are evaluated and their predictions combined via probability weighting:
\begin{equation}
    \hat{y}_m = \sum_{s=0}^{5} p_s\,\hat{y}_{m,s}.
    \label{eq:softroute}
\end{equation}
Stage~II is trained with MAE loss; the $[0,1]$ normalization ensures all six experts share a consistent prediction scale despite their widely varying age spans. For multi-modal subjects, per-modality predictions $\{\hat{y}_m\}_{m \in \mathcal{M}}$ are averaged at inference:
\begin{equation}
    \hat{y} = \frac{1}{|\mathcal{M}|}\sum_{m \in \mathcal{M}} \hat{y}_m.
    \label{eq:fusion}
\end{equation}
This prediction-level fusion allows any non-empty modality subset to be used at inference without retraining.

\section{Experiments}

\begin{figure}[t]
    \centering

    \includegraphics[width=\linewidth]{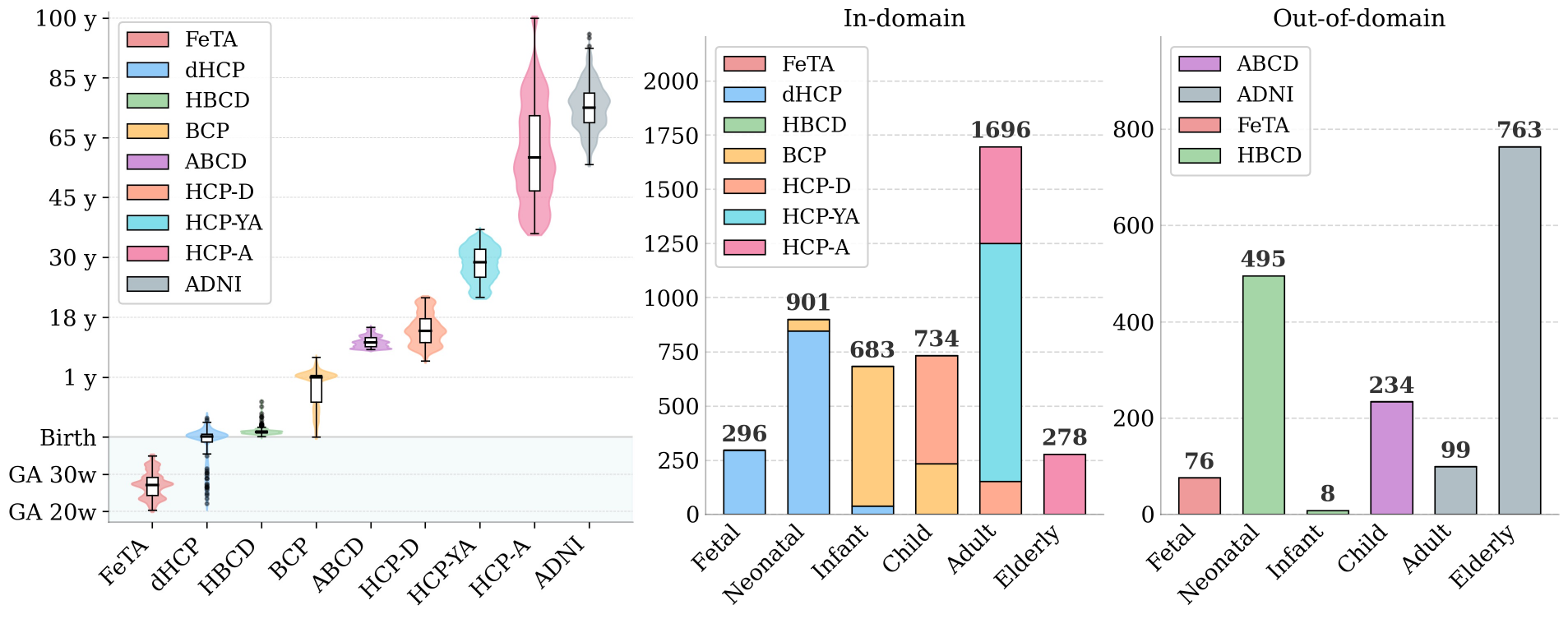}
        
    \caption{Nine neuroimaging datasets spanning gestational week~20 to age~100+. \textbf{Left}: age distributions per dataset; violin width encodes sample size ($\log_{10} N$), shaded region marks fetal subjects. \textbf{Right}: sessions per developmental stage, stacked by dataset.}
    \label{fig:distribution}
\end{figure}

\subsection{Experimental Setup}

We evaluate our framework on nine publicly available neuroimaging datasets spanning the full human lifespan, divided into in-domain and out-of-domain groups. Datasets were selected to cover all six developmental stages and three MRI modalities, ensuring comprehensive lifespan evaluation. All downstream data splits are performed at the subject level, with all sessions and modalities from each subject assigned to the same partition to avoid data leakage. More details about datasets can be found in Figure~\ref{fig:distribution} and Supplementary Section~A.

\paragraph{In-domain datasets} comprise five datasets used for MAE pretraining and downstream supervised training and validation.
\begin{itemize}
    \item \textbf{BCP: Baby Connectome Project}~\cite{howell2019unc} is a longitudinal study of brain development and connectivity from birth through early childhood.
    \item \textbf{dHCP: Developing Human Connectome Project}~\cite{eyre2021developing} is a longitudinal study mapping brain development in utero and through the neonatal period.
    \item \textbf{HCP-D: Human Connectome Project -- Development}~\cite{somerville2018lifespan} is a large-scale cross-sectional and longitudinal study of brain connectivity development in children, adolescents, and young adults aged 5--21.
    \item \textbf{HCP-YA: Human Connectome Project -- Young Adult}~\cite{van2013wu} is a large-scale study mapping brain circuits and their relationship to behavior and genetics in healthy young adults.
    \item \textbf{HCP-A: Human Connectome Project -- Aging}~\cite{bookheimer2019lifespan} is a large-scale cross-sectional study of brain aging in adults aged 36--100+ across four sites.
\end{itemize}

\paragraph{Out-of-domain datasets} comprise four datasets seen during unlabeled MAE pretraining but excluded from supervised training.
\begin{itemize}
    \item \textbf{ABCD: Adolescent Brain Cognitive Development}~\cite{casey2018adolescent} is a large multi-site longitudinal study of brain and cognitive development in children aged 9--15, spanning 21 US sites with diverse scanner protocols.
    \item \textbf{ADNI: Alzheimer's Disease Neuroimaging Initiative}~\cite{jack2008alzheimer} is a longitudinal multi-site study of Alzheimer's disease progression providing T1w-only data. We use 265 cognitively normal (CN) sessions for OOD evaluation; clinical groups are reserved for the brain age gap analysis in Section~\ref{sec:bag}.
    \item \textbf{FeTA: Fetal Tissue Annotation}~\cite{payette2021automatic} is a fetal MRI segmentation dataset providing T2w-only in utero scans spanning gestational weeks 20--35.
    \item \textbf{HBCD: Healthy Brain and Child Development}~\cite{volkow2024healthy} is a large nationwide US consortium study of early brain development providing T1w/T2w scans from birth to 31 postnatal weeks.
\end{itemize}

\subsection{Metrics}

For brain age regression, we report mean absolute error (MAE) and standard deviation (STD) in years, computed separately for each of the six developmental stages:
\begin{equation}
    \mathrm{MAE} = \frac{1}{N}\sum_{i=1}^{N}|\hat{y}_i - y_i|, \qquad
    \mathrm{STD} = \sqrt{\frac{1}{N}\sum_{i=1}^{N}\bigl(|\hat{y}_i - y_i| - \mathrm{MAE}\bigr)^2},
\end{equation}
where $\hat{y}_i$ and $y_i$ are the predicted and chronological ages of the $i$-th sample, and $N$ is the number of evaluation samples: sessions for Ours-M (after late fusion) and scans for all other methods (each scan evaluated independently). Stage-level reporting is essential because prediction difficulty varies substantially across the lifespan, with early stages spanning only weeks and adult stages spanning decades. For age stage classification, we report the confusion matrix $\mathbf{C} \in \mathbb{R}^{6\times6}$, where entry $C_{ij}$ denotes the number of samples with true stage $i$ predicted as stage $j$, and overall accuracy:
\begin{equation}
    \mathrm{Accuracy} = \frac{\sum_{i} C_{ii}}{\sum_{i,j} C_{ij}}.
\end{equation}

\subsection{Baselines}

We compare against three representative brain age prediction methods:
\begin{itemize}
    \item \textbf{SFCN}~\cite{peng2021accurate} is a lightweight seven-block fully convolutional network ($\sim$3M parameters) that treats age prediction as soft classification over discretized bins, training with KL divergence against Gaussian-smoothed age labels. Its compact design and strong inductive bias make it a widely adopted baseline in neuroimaging age estimation.
    \item \textbf{ORDER}~\cite{shah2024ordinal} reframes brain age prediction as ordinal classification with a ResNet backbone. A distance-regularized loss encourages the feature space to reflect ordinal age relationships by weighting pairwise feature distances proportionally to age differences between samples, directly addressing the regression-to-the-mean bias common in age estimation.
    \item \textbf{TSAN}~\cite{cheng2021brain} adopts a coarse-to-fine two-stage cascade: the first stage produces a rough age estimate, which the second stage refines using dense multi-scale feature connections. A ranking loss is applied jointly with MSE to explicitly preserve ordinal age relationships across subjects.
    \item \textbf{Ours-S} is a single-modal variant of our framework that processes each modality independently, used to isolate the contribution of multi-modal fusion from architectural gains.
\end{itemize}
All methods are trained on the same in-domain data and process each scan independently as a single-modal input without cross-modal fusion.

\subsection{Results}

\subsubsection{Stage~I: Age Stage Classification.}
\begin{figure}[tbp]
    \centering
    \includegraphics[width=\linewidth]{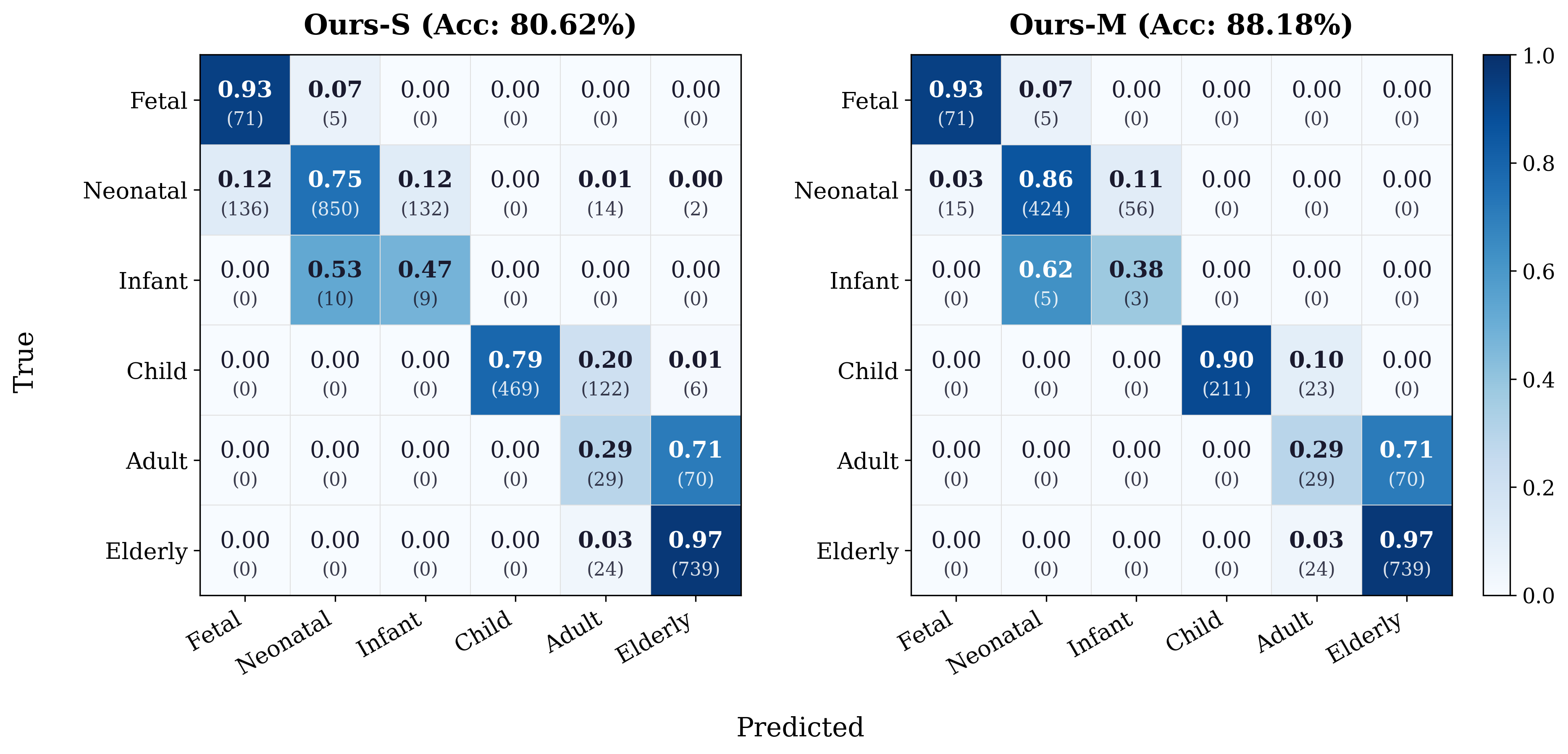}
    \caption{Stage~I confusion matrices on the out-of-domain datasets. Each cell reports the normalized fraction and raw count. Predicted stages are obtained via $\hat{s} = \arg\max_s p_s$, where $\mathbf{p} = \mathrm{softmax}(\hat{\mathbf{l}})$ is the Stage~I probability distribution; the same $\mathbf{p}$ is passed to Stage~II for soft expert weighting. \textbf{Left}: Ours-S evaluated per modality (2,688 predictions). \textbf{Right}: Ours-M evaluated per session after late fusion (1,675 predictions).}
    \label{fig:confusion}
\end{figure}
\paragraph{Overall Accuracy.}
Figure~\ref{fig:confusion} shows confusion matrices of the Stage~I classifier evaluated on the out-of-domain datasets. Since Ours-S processes each modality independently, it produces 2,688 predictions, whereas Ours-M fuses all available modalities per session to yield 1,675 session-level predictions. Ours-M achieves an overall accuracy of 88.18\%, outperforming Ours-S (80.62\%). Notably, Ours-M reduces cross-boundary errors: while Ours-S misclassifies some neonatal subjects as adult or elderly, Ours-M confines all errors to adjacent age groups, indicating that multi-modal aggregation produces more reliable stage predictions with fewer long-range classification errors. This behavior is important for Stage II regression, as adjacent-stage routing errors incur substantially smaller age penalties than long-range misclassifications. In addition, fetal and neonatal stages are classified with near-perfect accuracy by Ours-M, suggesting that the model handles visually distinct inter-stage boundaries reliably.

\paragraph{Adult Stage Boundary Analysis.}
Adult accuracy remains relatively low for both models (29\%), as the out-of-domain adult subjects come exclusively from ADNI, whose participants span ages 56--96---a range that straddles the adult/elderly boundary (65 years) defined during training. This mismatch makes adult--elderly confusion more likely, and is reflected in the large Adult out-of-domain MAE of 6.74 years in Table~\ref{tab:generalization}. Nevertheless, boundary misclassifications incur only modest regression error, as adult and elderly stages occupy \textit{adjacent} age ranges.

\subsubsection{Stage~II: Brain Age Regression.}
\begin{table}[tbp]
\centering
\caption{Brain age estimation results (MAE\,/\,STD in years, $\downarrow$). Best values in \textbf{bold}. Results are aggregated by developmental stage across all datasets using unified stage boundaries.}
\label{tab:generalization}
\resizebox{\columnwidth}{!}{%
\begin{tabular}{lcccccc}
\toprule
\textbf{Method} & \textbf{Fetal} & \textbf{Neonatal} & \textbf{Infant} & \textbf{Child} & \textbf{Adult} & \textbf{Elderly} \\
\midrule
\multicolumn{7}{l}{\textit{In-domain}} \\
\midrule
SFCN   & 0.87 / 0.10 & 0.60 / 0.06 & 0.44 / 0.32 & 0.96 / 1.24 & \textbf{2.93} / 3.95 & 6.96 / 7.04 \\
ORDER  & 0.21 / 0.07 & 0.05 / 0.05 & 0.32 / \textbf{0.21} & 1.28 / \textbf{1.15} & 3.82 / \textbf{3.82} & 8.92 / 8.14 \\
TSAN   & 0.14 / 0.40 & 0.12 / 0.16 & 0.65 / 1.24 & 1.97 / 2.45 & 3.71 / 4.83 & 6.30 / 8.36 \\
Ours-S & \textbf{0.01} / 0.02 & \textbf{0.02} / \textbf{0.03} & 0.15 / \textbf{0.21} & 1.20 / 1.63 & 3.67 / 4.90 & 5.50 / 6.82 \\
Ours-M & \textbf{0.01} / \textbf{0.01} & \textbf{0.02} / \textbf{0.03} & \textbf{0.13} / 0.22 & \textbf{0.91} / 1.21 & 3.16 / 4.17 & \textbf{4.92} / \textbf{6.16} \\
\midrule
\multicolumn{7}{l}{\textit{Out-of-domain}} \\
\midrule
SFCN   & 0.97 / 0.12 & 0.80 / 0.49 & 0.58 / 0.34 & 5.36 / 4.19 & 31.66 / 6.53 & 44.62 / 11.57 \\
ORDER  & 0.25 / \textbf{0.07} & 0.16 / 0.25 & 0.44 / \textbf{0.13} & 7.56 / \textbf{3.78} & 26.00 / 9.30 & 36.49 / 21.75 \\
TSAN   & 0.69 / 1.15 & 0.32 / 1.08 & 1.22 / 3.01 & 4.68 / 4.57 & 47.74 / \textbf{5.33} & 57.85 / 14.41 \\
Ours-S & \textbf{0.04} / \textbf{0.07} & \textbf{0.09} / \textbf{0.16} & \textbf{0.19} / 0.16 & 6.26 / 11.72 & \textbf{6.74} / 5.53 & \textbf{5.36} / \textbf{6.88} \\
Ours-M & \textbf{0.04} / \textbf{0.07} & \textbf{0.09} / \textbf{0.16} & \textbf{0.19} / 0.17 & \textbf{3.88} / 7.94 & \textbf{6.74} / 5.53 & \textbf{5.36} / \textbf{6.88} \\
\bottomrule
\end{tabular}%
}
\end{table}
\paragraph{In-Domain Performance.}
As shown in Table~\ref{tab:generalization}, Ours-M achieves competitive in-domain performance across developmental stages. Prediction errors for fetal and neonatal stages approach zero (MAE $\leq$ 0.02 years). For the child stage, Ours-M achieves the lowest MAE of any method (0.91 years). For the adult stage, it achieves an MAE of 3.16 years, outperforming all single-stage methods except SFCN. For the elderly stage, Ours-M achieves the lowest MAE of any method (4.92 years). In addition, multi-modal fusion provides a consistent gain over its single-modal counterpart Ours-S, with 13\% lower average in-domain MAE.

\paragraph{Out-of-Domain Generalization.}
On out-of-domain data, baselines exhibit severe degradation, with adult and elderly MAE exceeding 26 and 36 years, respectively. This degradation likely stems from single-stage global regression over the full lifespan: without stage-conditioned routing, predictions become highly sensitive to domain shift, allowing errors to span the entire age axis. Our Stage~I classifier addresses this by routing each subject to a Stage~II expert calibrated to a narrower developmental window. Ours-M reduces elderly MAE to 5.36 years, yielding an 85\% reduction over the best baseline. Multi-modal fusion further improves out-of-domain robustness by 13\% over Ours-S. However, the advantage of multi-modal fusion is not observed uniformly across all stages, as some out-of-domain datasets contain only a single modality (e.g., ADNI: T1w; FeTA: T2w).

\paragraph{Per-Dataset Analysis.}
Per-dataset breakdowns and a detailed analysis of the anomalous baseline failures on ADNI are provided in Supplementary Section~B.

\subsubsection{Brain Age Gap in Clinical Populations.}
\label{sec:bag}
\paragraph{Setup.}
The brain age gap (BAG), defined as $\mathrm{BAG} = \hat{y} - y$, quantifies the deviation of predicted brain age from chronological age, with a positive value indicating accelerated brain aging~\cite{baecker2021machine}.
We evaluate BAG on ADNI~\cite{jack2008alzheimer} subjects stratified by diagnostic group along the Alzheimer's disease continuum: cognitively normal (CN), subjective memory concern (SMC, subjects reporting memory difficulties without objective impairment), early and late mild cognitive impairment (EMCI, LMCI), a general MCI category, and Alzheimer's disease (AD).
This sequence represents a clinically established progression from healthy aging through prodromal stages to dementia.

\paragraph{Results.}
As shown in Table~\ref{tab:bag}, mean BAG shows an overall upward trend along the disease continuum. CN subjects show a modest positive BAG ($+1.24$\,y), while AD patients exhibit substantially accelerated brain aging ($+7.74$\,y).
Intermediate groups also show elevated mean BAG: SMC ($+4.67$\,y), EMCI ($+4.11$\,y), LMCI ($+6.32$\,y).
The MCI group ($+3.08$\,y) falls slightly below EMCI, consistent with MCI being a heterogeneous category spanning early and late impairment in the ADNI cohort. These results suggest an association between predicted brain age and clinical status across the AD spectrum.

\begin{table}[tbp]
\centering
\caption{Brain age gap (BAG\,$=$\,$\hat{y}-y$, years) by ADNI diagnostic group along the Alzheimer's disease continuum. Positive BAG indicates accelerated brain aging.}
\label{tab:bag}
\setlength{\tabcolsep}{4pt}
\begin{tabular}{lcccc c}
\toprule
\textbf{Group} & \textbf{$N$} & \textbf{Mean $y$} & \textbf{Mean $\hat{y}$} & \textbf{Mean BAG} & \textbf{STD} \\
\midrule
CN   & 265 & $76.90$ & $78.14$ & $+1.24$ & $6.93$ \\
SMC  &  21 & $74.10$ & $78.76$ & $+4.67$ & $4.79$ \\
EMCI & 236 & $72.91$ & $77.02$ & $+4.11$ & $7.91$ \\
MCI  & 102 & $77.08$ & $80.16$ & $+3.08$ & $6.98$ \\
LMCI & 141 & $73.28$ & $79.59$ & $+6.32$ & $8.68$ \\
AD   &  97 & $76.05$ & $83.79$ & $+7.74$ & $7.32$ \\
\bottomrule
\end{tabular}
\end{table}

\begin{table}[b]
\centering
\caption{Ablation study results (MAE / STD in years, $\downarrow$, in-domain val set). Best values in \textbf{bold}. Additional per-modality ablation results are provided in Supplementary Section C.}
\label{tab:ablation}
\resizebox{\columnwidth}{!}{%
\begin{tabular}{lcccccc}
\toprule
\textbf{Method} & \textbf{Fetal} & \textbf{Neonatal} & \textbf{Infant} & \textbf{Child} & \textbf{Adult} & \textbf{Elderly} \\
\midrule
Ours-M        & \textbf{0.01} / \textbf{0.01} & \textbf{0.02} / \textbf{0.03} & 0.13 / 0.22 & \textbf{0.91} / \textbf{1.21} & \textbf{3.16} / \textbf{4.17} & \textbf{4.92} / \textbf{6.16} \\
\midrule
w/o Mod MoE   & \textbf{0.01} / \textbf{0.01} & \textbf{0.02} / \textbf{0.02} & 0.18 / 0.20 & 1.16 / 1.65 & 3.86 / 5.04 & 5.06 / 6.68 \\
w/o Stage MoE & 0.08 / 0.03 & 0.07 / 0.07 & 0.22 / 0.20 & 1.27 / 1.65 & 3.49 / 4.64 & 5.02 / 6.63 \\
w/o MAE Pre.  & \textbf{0.01} / 0.02 & \textbf{0.02} / \textbf{0.02} & \textbf{0.11} / \textbf{0.15} & 1.17 / 1.63 & 3.97 / 5.86 & 5.77 / 7.31 \\
w/o Stage~I   & 0.08 / 0.03 & 0.12 / 0.10 & 0.21 / 0.28 & 1.22 / 1.66 & 3.40 / 4.55 & 5.51 / 6.93 \\
Encoder Frozen & \textbf{0.01} / 0.02 & \textbf{0.02} / \textbf{0.03} & 0.18 / 0.25 & 1.36 / 1.76 & 3.87 / 5.23 & 5.89 / 7.23 \\
\bottomrule
\end{tabular}%
}
\end{table}

\subsection{Ablation Study}

\paragraph{Setup.}
To validate the key design choices in our framework, we evaluate five ablation variants against the full model, each isolating a single component while keeping the rest fixed.
\textbf{w/o Mod MoE} removes the Modality MoE from the encoder, replacing expert routing with a shared convolutional representation for all modalities.
\textbf{w/o Stage MoE} retains Stage~I but replaces the six per-stage regression experts in Stage~II with a single shared head, removing the stage-conditioned probability weighting.
\textbf{w/o MAE Pre.} trains the encoder from random initialization, removing the MAE-pretrained weights that otherwise provide a general-purpose feature prior before supervised fine-tuning.
\textbf{w/o Stage~I} eliminates the coarse classification stage entirely, replacing the two-stage hierarchy with a single global regression head that predicts age directly from encoder features.
\textbf{Encoder Frozen} keeps the full architecture but freezes the encoder during Stage~II training, preventing any update to the encoder weights while the regression heads are optimized.

\paragraph{Results.}
As shown in Table~\ref{tab:ablation}, removing individual components generally degrades performance across multiple developmental stages, supporting the effectiveness of the proposed design.

\section{Conclusion}

We presented a two-stage multi-modal framework for lifespan brain age prediction, integrating T1w, T2w, and FA MRI via late fusion to accommodate missing modalities. Evaluated on nine datasets spanning fetal to elderly stages, our method reduces MAE by 13\% in-domain and 78\% out-of-domain compared to the best baseline, with an additional 12--13\% gain from multi-modal fusion over the single-modal variant. Ablation studies further validate the contribution of stage-conditioned routing, modality-specific experts, and MAE pretraining.

\begin{credits}
\subsubsection{\ackname}
{\emergencystretch=2em
This work was supported by NIH grants R00HD103912, R01MH133313 (Y.W.).
\par}

\subsubsection{\discintname}
The authors declare that they have no competing interests relevant to this work.
\end{credits}

\bibliographystyle{splncs04}
\bibliography{references_paper}

\end{document}